%% file: ms.tex
\documentclass[11pt]{article}
\input{header}

\usepackage{fullpage, parskip} 
\usepackage{cleveref}
\usepackage{booktabs}

\definecolor{RevisedColor}{RGB}{255, 0, 0}
\newcommand{\edit}[1]{#1}

\hypersetup{pdfborder = {0 0 0.5 [3 3]}, colorlinks = true, linkcolor = BrightRed, citecolor = SkyBlue}

\title{Quantifying the limits of human athletic performance: A Bayesian analysis of elite decathletes}
\author{Paul-Hieu V.~Nguyen\thanks{Department of Statistics, University of Wisconsin--Madison. \url{pvnguyen5@wisc.edu}}  \and James M. Smoliga \thanks{Department of Rehabilitation Science, Tufts University. \url{james.smoliga@tufts.edu}} \and Benton Lindaman \thanks{Department of Rehabilitation Science, Tufts University. \url{benton.lindaman@tufts.edu}} \and Sameer K. Deshpande \thanks{Department of Statistics, University of Wisconsin--Madison. \url{sameer.deshpande@wisc.edu}}
}

\begin{document}
\maketitle

\begin{abstract}
\input{abstract}
\end{abstract}

\section{Introduction}
\label{sec:intro}
\input{intro}



\section{Data and Model}
\label{sec:model}
\input{model}

\section{Posterior inference and simulated performance}
\label{sec:experiments}
\input{experiments}


\section{Discussion}
\label{sec:discussion}
\input{discussion}

\section*{Acknowledgements}
\label{sec:acknowledgements}
\input{acknowledgement}

\bibliographystyle{apalike}
\bibliography{decathlon_refs}

\appendix
\renewcommand{\theequation}{\thesection\arabic{equation}}
\renewcommand{\thefigure}{\thesection\arabic{figure}}  
\renewcommand{\thetable}{\thesection\arabic{table}} 

\setcounter{equation}{0}
\setcounter{section}{0}
\setcounter{subsection}{0}
\setcounter{subsubsection}{0}

\section{Additional results}
\label{app:appendix}
\input{appendix}

\end{document}

%% file: header.tex
\usepackage{amssymb, amsthm}
\usepackage{algorithm}
\usepackage{algorithmicx}
\usepackage{amsmath, amssymb, amsthm}
\usepackage{float,multirow,subcaption, setspace}
\usepackage{graphicx}
\usepackage{comment}
\usepackage{enumitem}
\usepackage{tikz}
\usepackage{bm, bbm}
\usetikzlibrary{shapes.geometric, positioning}
\usetikzlibrary{quotes, angles}
\usepackage{rotating}
\usepackage{hyperref}
\usepackage{natbib}
\usepackage{array}

\definecolor{SkyBlue}{RGB}{14, 118, 188}
\definecolor{BrightRed}{RGB}{223, 82, 78}
\definecolor{BurntOrange}{HTML}{BF5900}

\newcommand{\normaldist}[2]{\mathcal{N}\left(#1,#2\right)} 
\newcommand{\igammadist}[2]{\textrm{Inv.~Gamma}\left(#1,#2\right)} 

\theoremstyle{plain}

\theoremstyle{definition}

\theoremstyle{plain}

%% file: abstract.tex
Because the decathlon tests many facets of athleticism, including sprinting, throwing, jumping, and endurance, many consider it to be the ultimate test of athletic ability.
On this view, estimating the maximal decathlon score and understanding what it would take to achieve that score provides insight into the upper limits of human athletic potential.
To this end, we develop a Bayesian composition model for forecasting how individual \edit{decathletes} perform in each of the 10 decathlon events of time.
Besides capturing potential non-linear temporal trends in performance, our model carefully captures the dependence between performance in an event and all preceding events. 
Using our model, we can simulate and evaluate the distribution of the maximal possible scores and identify profiles of \edit{decathletes} who could realistically attain scores approaching this limit. 

%% file: intro.tex
\subsection{Motivation: realistic max decathlon score}

The decathlon is a combined track-and-field event consisting of ten disciplines spread over two days.  
These events test multiple facets of athletic ability, including sprinting, jumping, throwing, technique, and endurance. The decathlon is widely regarded as the ultimate test of athletic ability due to its diversity of events and demand on both the body and mind \edit{\citep{edouard2010causes}}.
The order of events is consistent for each decathlon. 
\edit{Though technique is valuable for each discipline, we note that generally, d}ay one emphasizes explosiveness --- \edit{decathlete}s compete in the 100m, long jump (LJ), shot put (SP), high jump (HJ), and 400m --- while day two focuses on technique and endurance, featuring the 110m hurdles, discus throw (DT), pole vault (PV), javelin throw (JT), and 1500m. 
Decathletes earn points based on their performance in each discipline based on a scoring table developed by the World Athletics.
Their overall decathlon score is the sum of each event's points. 
The scoring system is \edit{designed to balance high-level performances with overall competence. Excelling in only a single discipline will produce exponentially increasing rewards for that event, while underperformance in other events diminishes the overall score.} 

More specifically, let $Y_{e}$ measure \edit{a} decathlete's performance in event $e.$
For track events (i.e., the 100m, 400m, 110m hurdles, or 1500m), $Y_{e}$ is a time and smaller values of $Y_{e}$ correspond to better performance.
For all others (i.e., LJ, SP, HG, DT, PV, and JT), $Y_{e}$ is distance and larger values correspond to better performance.
Decathletes earn $a_{e} - (b_{e} - Y_{e})^{c_{e}}$ points for track events and $a_{e} - (Y_{e} - b_{e})^{c_{e}}$ points for all other events, where $a_{e}, b_{e},$ and $c_{e}$ are event-specific coefficients (see \Cref{tab:point_params}).

\begin{figure}[h]
  \centering
    \includegraphics[width=.9\linewidth]{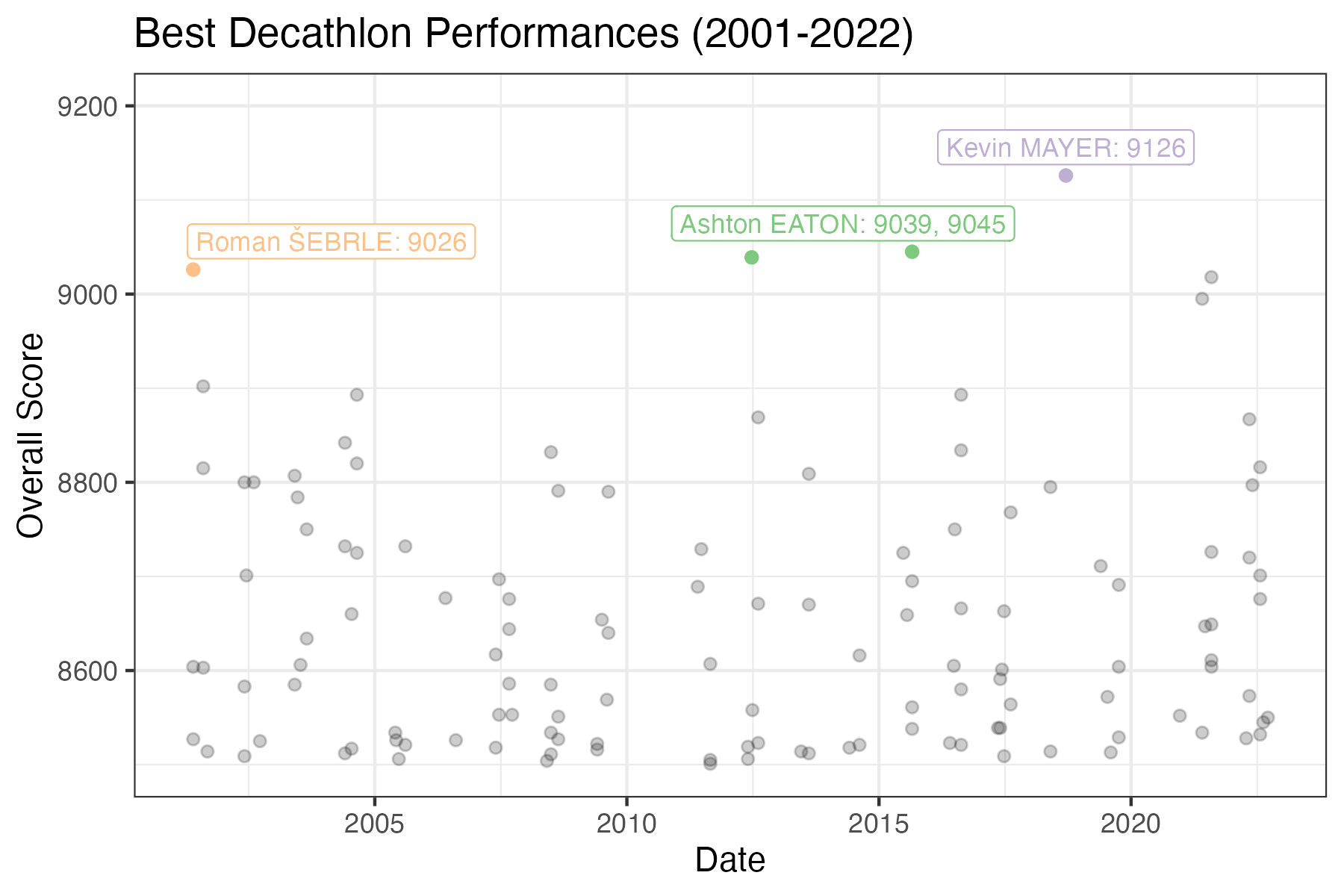}
    \caption{Top decathlon scores from 2001-2022. We highlight the previous world record holders and scores. All decathlon performances greater than 8500 are included in this graph (n = 126).}
    \label{fig:best_scores}
\end{figure}

\Cref{fig:best_scores} shows the best decathlon scores from 2001 to 2022, with world records by Roman \v{S}erble (2001), Ashton Eaton (2012 and 2015), and Kevin Mayer (2019). 
These \edit{decathlete}s differ in their event-specific strengths, with \v{S}erble excelling in jumping events \edit{and javelin throw}, Eaton excelling in sprinting, and Mayer doing well across all disciplines. 
Mayer broke the previous world record by roughly eighty points. 
Given these trends, we ask whether a new performance threshold, say another eighty point increase for a 9200 point total, is realistically attainable, and what combination of event abilities would be required to achieve 9200 points.

Based on World Athletics scoring table, if a decathlete managed to attain world-record performance in each individual event, their score would be 12,676. 
Of course, it is unrealistic to expect \edit{a decathlete} to perform at a world-record level in all ten disciplines.
If, on the other hand, a decathlete managed to attain the highest score in each event \emph{that has ever been observed specifically in a decathlon}, their score would be 10,669. 
Even this hypothetical is unrealistic as it requires a decathlete to attain peak performance in all disciplines at exactly the same time.
Intuitively, we would expect individual event performance to vary with age \citep{villaroel2011elite}, with performance peaking in different events at different times.

Motivated by this, we build age curves for individual decathlete's performances in each discipline.
By applying World Athletics' scoring table to forecasted individual event performances, we can predict an individual decathlete's overall score over the course of their career. 
To capture the inter-event dependencies, we specifically fit \emph{compositional} age curves in which performance in one discipline depends not only on a non-linear function of age but also on performance in the immediately preceding disciplines. 
That is, our model for LJ performance depends on the observed 100m.
We take a Bayesian approach, which allows us to quantify and propagate uncertainty about forecasted discipline performance through to final decathlon score in a coherent fashion. 

\subsection{Related Literature}
\label{subsec:background}

The majority of prior research on the decathlon has been descriptive in nature focusing on clustering different disciplines \citep{cox2002analysis,woolf2007grouping,walker2015performance,schomaker2011model} and creating archetypes of decathletes who perform similarly across different discipline clusters \citep{dziadek2022principal,kenny2005determinants,van2002performance}. 
\citet{cox2002analysis}, for instance, applied \edit{multiple clustering and dimension scaling methods} to group decathlon events, broadly separating the track disciplines \edit{and JT} from the field disciplines.
\edit{They noted that the HJ, PV, JT, and 1500m were not as consistent with cluster membership compared to other events}. 
\citet{woolf2007grouping} used cluster analysis based on personal best performances of elite decathletes and suggest track athletes may have a scoring advantage. 
\edit{\citet{schomaker2011model} analyzed data from the 2004 Olympics using factor analysis and identified three main groups, one each for speed, strength, and endurance events.}
Other authors have analyzed decathlon performance using principle component analysis (PCA). 
\citet{park2011multivariate} identified latent structures across events while \citet{dziadek2022principal} tracked how the structure of the decathlon shifts over \edit{a decathlete}'s career and found that \edit{decathlete}s broadly shifted from generalists to specializing in particular events. 
\edit{This finding may be a result of data availability across different stages of different \edit{decathlete}s' careers}.

Several authors have attempted to identify trade-offs between specializing in different disciplines.
\citet{van2002performance} found \edit{some} evidence of antagonistic traits, as well as tradeoffs between specialist and generalist phenotypes. 
\citet{aoki2015relationships} found physiological differences between athletes specializing in sprinting and jumping, though their analysis is not directly focused on the decathlon. 
\citet{walker2015performance,kenny2005determinants} argue that there is no evidence of event trade-offs when looking at the subpopulation of elite decathlon performers: top athletes perform uniformly well across all events, and they found positive correlation between all decathlon events. \edit{\citet{park2011multivariate} comment that the outcomes of these experiments differ between methods and sample, and their analysis agrees with \citet{kenny2005determinants}.}

Fewer studies have explored the decathlon from a predictive standpoint. 
\citet{battles2025predicting} use gamma regression to model \edit{a decathlete}'s career best using early-career decathlon performances. 
They found that the results from PV, JT, LJ, and SP were especially predictive of future performance.  
To the best of our knowledge, \citet{wimmer2011exploring} is the only other Bayesian analysis of decathlon data.
They fit semi-parametric latent variable models to cluster disciplines and model the effects of age, season, and year on decathlon results.

Age-performance relationships have also been explored outside of the decathlon. 
Researchers have sought to identify ages for peak performance in hockey \citep{schuckers2023estimation, brander2014estimating}, baseball \citep{fair2008estimated}, golf \citep{baker2023flexible}, and the triathlon \citep{villaroel2011elite}. 
\citet{griffin2022bayesian} use a Bayesian analysis to model individual sprinting and weightlifting events.


\subsection{Our contributions}
\label{sec:contributions}

We introduce a compositional Bayesian model to account for the multi-event, dependent nature of the decathlon. 
We model \edit{a decathlete}'s scores in an event as a function of the \edit{decathlete}'s age, their preceding-event scores within the same decathlon, and \edit{decathlete}-specific random intercepts.
We compare multiple models, with varying levels of flexibility and granularity, in an extensive set of experiments.
We show that compositional models accurately predict decathlon scores and allow for greater interpretability, enabling researchers to study relationships between events, than simpler models.
Using our probabilistic models, we obtain personalized decathlon and event-specific age curves to develop training programs, set goals for competitions, and model potential for future success. 
We further develop several real and synthetic \edit{decathlete} profiles, based on latent abilities in each individual decathlon event, and we simulate decathlon performances to investigate the distribution of decathlon scores from these \edit{decathlete} profiles. 
Through these simulations, we show that breaking the 9200 point threshold is unlikely, but still possible.


%% file: model.tex
\subsection{Data}

Our dataset contains the overall score and the results from each individual discipline from all completed decathlons between 2001 and 2022.
These data were initially provided by World Athletics and were later analyzed and distributed by \citet{battles2025predicting}\footnote{The data is available at \url{https://github.com/Battles186/DecathlonCareerBest.git}}.
The data contains observations from all decathlon performances with overall scores greater than 7000 for 2001-2008, greater than 6600 for 2009, and above 6400 for the years 2010-2022. 
We further truncated the data to include performances from only those decathletes who scored 6400 points or more at least four times.
This threshold is somewhat lower than \citet{battles2025predicting}, who kept only those decathletes who completed a decathlon in at least four seasons, and allows us to model a wider range of development trajectories. \edit{We note that previous decathlon analyses' results have differed depending on the method and sample \citep{park2011multivariate,van2002performance,kenny2005determinants}.}
We standardized each individual discipline's performance to have a mean of zero and standard deviation of one. 
Our final dataset includes 8668 decathlon performances from 1007 unique decathletes.

\subsection{Modeling decathlon performance over time}
To predict how a decathlete's overall scores evolve over the course of their career and to account for the fact that we have multiple observations per decathlete, we can fit a simple age curve with decathlete-specific random intercepts.
Letting $P_{i,j}$ be the total score earned by decathlete $i$ in decathlon $j$, a simple starting model asserts

\begin{equation}
\label{eq:baseline_model}
P_{i,j} = \tilde{\alpha}_{i} + \sum_{d}\tilde{\beta}_{d} \cdot \phi_{d}(\textrm{age}_{i,j})  + \epsilon_{i,j}, \quad \epsilon_{i,j} \sim \normaldist{0}{\sigma},
\end{equation}
where $\tilde{\alpha}_{i}$ is \edit{a} decathlete-specific random intercept, $\{\phi_{1}(\cdot), \ldots, \phi_{D}(\cdot)\}$ is some pre-specified basis of non-linear functions of age, and $\textrm{age}_{i,j}$ is the age of decathlete $i$ when they completed decathlon $j$.
To fit this model, we take a Bayesian approach based on the following relatively weakly informative priors, which we specify \emph{after} standardizing the observed $P_{i,j}$ values to have mean zero and variance one:
\begin{align*}
\begin{split}
    \tilde{\alpha}_{i} &\sim \normaldist{\mu_{\tilde{\alpha}}}{\sigma_{\tilde{\alpha}}}\\
    \mu_{\tilde{\alpha}} &\sim \normaldist{0}{1} \\
    \sigma_{\tilde{\alpha}}^2 &\sim \igammadist{2}{1} \\
    \tilde{\beta}_{d} &\sim \normaldist{{0}}{1} \\
    \tilde{\sigma}^2 &\sim \igammadist{2}{1}
\end{split}
\end{align*}

While it is relatively straightforward to fit the age curves in \Cref{eq:baseline_model} and to use them to identify when a given decathlete's performance will peak, simply modeling the overall points provides no insight into \emph{how} a given decathlete can obtain that peak.
That is, with such a simple model, it is impossible to determine whether a decathlete's forecasted performance is due to improved performance in any particular discipline. 
So, we instead model the outcomes of the individual events.
Specifically, let $Y_{i,j,e}$ record the performance of decathlete $i$ in discipline (hereafter ``event'') $e$ in decathlon $j.$
Then a natural model for individual event performances asserts for each $i, j$ and $e$ that
\begin{equation}
\label{eq:simple_model}
Y_{i,j,e} = \alpha_{i,e} + \sum_{d=1}^{D}{\beta_{d,e}\phi_{d}(\textrm{age}_{i,j})} + \epsilon_{i,j,e}; \quad \epsilon_{i,j,e} \sim \normaldist{0}{\sigma_e}, 
\end{equation}
where the $\alpha_{i,e}$'s are random intercepts specific to each combination of \edit{decathlete} and event. 
Because the model in \Cref{eq:simple_model} features event-specific basis coefficients $\beta_{e,d},$ it is flexible enough to allow the temporal evolution of individual event performances to vary across events.
In other words, the model in \Cref{eq:simple_model} allows for decathletes to obtain peak performance in different disciplines at different times.

To fit the model in \Cref{eq:simple_model}, we specify the priors
\begin{align*}
\begin{split}
        \label{eq:simple_priors}
        \alpha_{i,e} &\sim \normaldist{\mu_\alpha}{\sigma_\alpha}\\
        \mu_{\alpha} &\sim \normaldist{0}{1} \\
        \sigma^2_\alpha &\sim \igammadist{2}{1} \\
        \beta_{d,e} &\sim \normaldist{{0}}{1} \\
        \sigma^2_e &\sim \igammadist{2}{1}
\end{split}
\end{align*}

Despite its intuitive appeal, the model in \Cref{eq:simple_model} implicitly assumes that performance in each event is independent of performance in all other events.
Given the sequential nature of the decathlon --- namely, the events are run in the same, fixed order --- it is natural to suspect that individual event performances are not independent due to factors like fatigue and event similarity.
For instance, \edit{a} decathlete's performance in the 100m may impact their performance in the 400m, which takes place on the same day. \edit{\citet{beaulieu1995blood} found that \edit{decathlete}s' blood lactate levels differed before each event and were particularly high before the 400m.} 
To better capture inter-event dependencies, we propose a \emph{compositional} elaboration of the model in \Cref{eq:simple_model} in which performance in each event depends on performance in all preceding events:
\begin{equation}
\label{eq:compositional_model}
Y_{i, j, e} = \alpha_{i,e} + \sum_{d=1}^{D}{\beta_{d,e}\phi_{d}(\textrm{age}_{i,j})} + \sum_{m = 1}^{e-1}{\gamma_{m,e} Y_{i,j,m}}  + \epsilon_{i,j,e}; \quad \epsilon_{i,j,e} \sim \normaldist{0}{\sigma_{e}}. 
\end{equation}
To illustrate the difference between the models in \Cref{eq:simple_model} and \Cref{eq:compositional_model}, consider modeling shot put (SP) performance.
Whereas the simple model in \Cref{eq:simple_model} only uses decathlete age and identity to model SP performance, the composition model in \Cref{eq:compositional_model} additionally accounts for performance in the long jump (LJ) and 100m, which take place immediately before SP. 
In this context, the coefficient $\gamma_{100m,SP}$ captures, up to scaling, the conditional correlation between 100m performance and SP performance holding age and LJ performance fixed.
If there is a positive (resp.\ negative) conditional correlation --- that is, if running a faster 100m is associated with achieving longer (resp.\ shorter) shot put distances --- we would expect $\gamma_{100m,SP}$ to be negative (resp.\ positive).
On the other hand, if there is essentially no conditional dependence between sprinting and shot putting abilities, we would expect $\gamma_{SP,100m}$ to be close to zero. 
\edit{This is an alternate representation of the grouped structure previously explored within the decathlon literature by \citet{schomaker2011model,woolf2007grouping,cox2002analysis}. Our approach reflects how the decathlon is physically performed, whereas other types of analyses do not account for the sequential nature of the competition.}

We specify analogous priors for the parameters in \Cref{eq:compositional_model} as we did for \Cref{eq:simple_model}:
\begin{align*}
\begin{split}
\alpha_{i,e} &\sim \normaldist{\mu_\alpha}{\sigma_\alpha}\\
\mu_{\alpha} &\sim \normaldist{0}{1} \\
\sigma^2_\alpha &\sim \igammadist{2}{1} \\
\beta_{d,e} &\sim \normaldist{{0}}{1} \\
\gamma_{m,e} &\sim \normaldist{{0}}{1} \\ 
\sigma^2_e &\sim \igammadist{2}{1}
\end{split}
\end{align*}

%% file: experiments.tex
\subsection{Model comparison and validation}
\label{subsec:model_comp}

\textbf{Model comparison}.
We fit each model using the \textbf{rstan} \citep{rstan_package} interface to \textsf{Stan}.
We used the package defaults, simulating four Markov chains for 2,000 iterations each and discarding the first 1,000 iterates as ``burn-in.''
We performed a 10-fold cross-validation study to compare the predictive performance of two versions of each model, one that set $\{\phi_{d}\}$ to be a cubic polynomial basis and one that set $\{\phi_{d}\}$ to be a cubic splines basis with interior knots at age deciles. 

We compared the out-of-sample predictive accuracy on real decathlon data for each version of each model in two ways.
First, which we call the ``general'' case, we created ten 90\%-10\% training/testing splits where individual decathlete observations were randomly held out in the test set.
In the second framework, which we call the ``tail'' case, we held out just the last observed decathlon for a random 10\% of the decathletes. 
Generally speaking, all models achieved very similar out-of-sample mean square errors in both cases; see \Cref{tab:gen_table,tab:future_table} for full tabulations.
In the general case, the average out-of-sample standardized mean square error 
(SMSE\footnote{Standardized mean square error is the mean square error divided by the variance of the \edit{training set responses. A model with worse predictive performanance than another model will have a larger SMSE. 
A model with perfect out-of-sample accuracy will have an SMSE of 0, whereas a baseline naive model that simply predicts the average response will have an SMSE of 1.}}) 
for the baseline points model in \Cref{eq:baseline_model} was 0.234 and 0.235 using the cubic polynomial and cubic splines basis.
Both versions of the simple (\Cref{eq:simple_model}) and compositional (\Cref{eq:compositional_model}) models achieved mean out-of-sample SMSEs of 0.235.
In the ``tail'' case, we found that the baseline points model with cubic polynomial basis had ever-so-slightly smaller SMSE than the compositional model with the same non-linearities (0.358 vs 0.362). 
In this setting, we found that modeling the age-event relationship with a cubic polynomial tends to perform slightly better than modeling with a spline. 

Although our compositional model with a cubic polynomial basis had slightly worse predictive performance than the baseline model for overall points, it provides much more insight into the underlying inter-event dynamics.
To support this choice further, we conducted two additional simulation studies that verify the ability of the compositional model to detect inter-event dependencies when present.

\textbf{Ability to recover model parameters.} 
First, we demonstrate that the compositional model was powered to recover the model parameters.
At a high-level, we fixed values of all parameters in \Cref{eq:compositional_model}, generated 200 synthetic datasets of the same size as our decathlon dataset, fit our model to those synthetic datasets, and assessed how well we estimated the data-generating parameter values.
We specifically assessed the extent to which the 95\% posterior credible intervals for each parameter covered the true data-generating parameter values.
Generally speaking, with the \edit{single exception} of \edit{the coefficient} for JT \edit{when modeling} PV (90.5\%), the vast majority of the intervals displayed near-nominal 95\% coverage; see \Cref{tab:prop_table_cube_comp} for a full tabulation.
Overall, the high coverage across virtually all predictors suggests that the model is well-calibrated for recovering inter-event relationships.

\begin{figure}[ht]
    \centering
        \includegraphics[width = .9\linewidth]{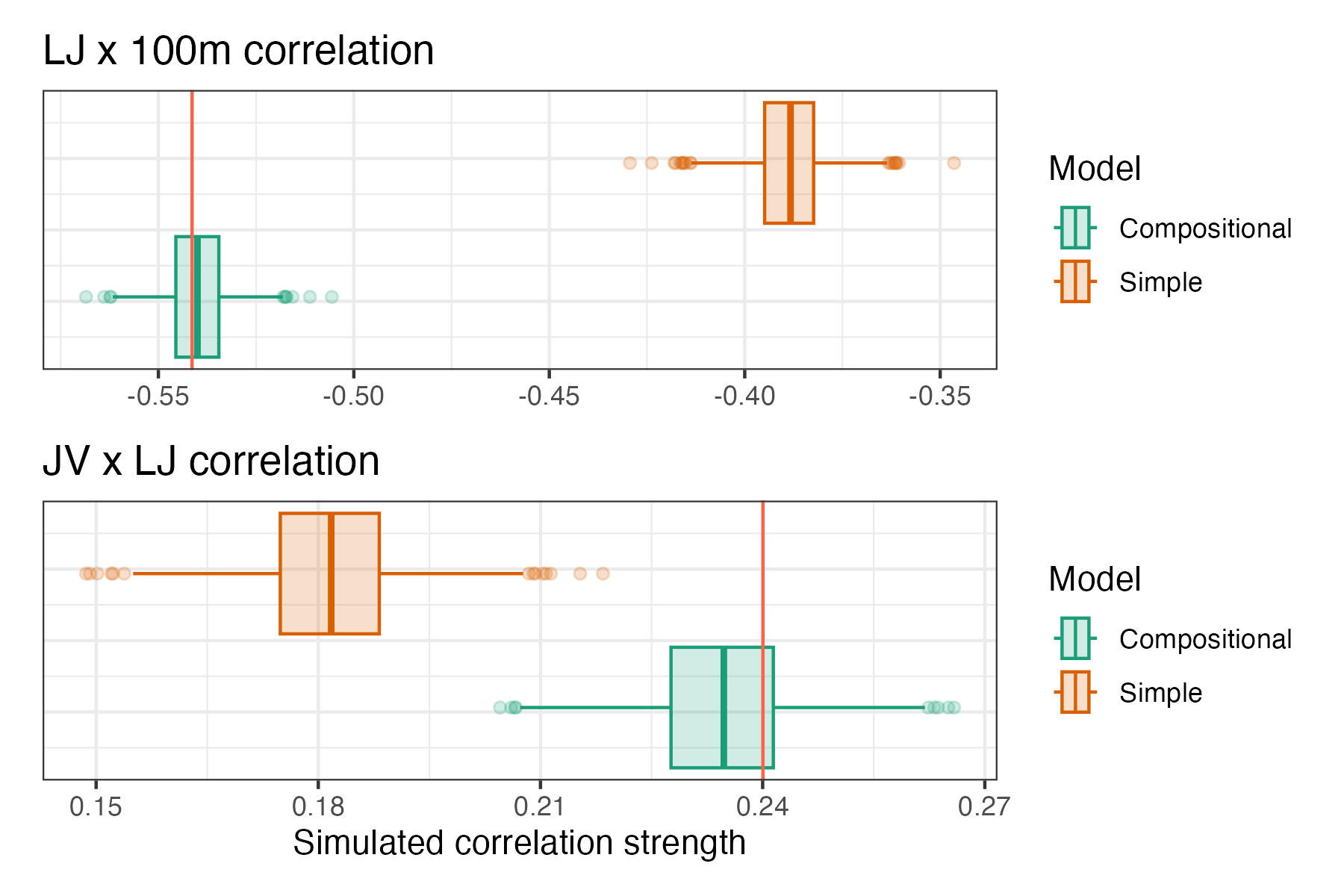} 
    \caption{Boxplots depicting the posterior predictive correlations for the simple and compositional models over 2000 simulated datasets between the 100m and long jump (top) and javelin and long jump (bottom). For each model, the 25th, 50th, and 75th percentile is marked in the boxplot. The empirical correlation from the observed data is marked with a red line.}
    \label{fig:cor_box_plots}
\end{figure}

\textbf{Posterior predictive checks.}
Once we fit our simple and compositional model with cubic polynomial basis, we generated 2,000 decathlon datasets from the posterior predictive distribution using the same \edit{decathlete}s and ages as in the observed data. 
Using these simulated datasets, we computed the posterior predictive correlation between every pair of individual events.
\edit{The colored boxplots in }\Cref{fig:cor_box_plots} show the posterior predictive distribution of these correlations for selected pairs of events for both models\edit{, with the true observed correlation represented with a red line}.
We see that the true observed inter-event correlation\edit{s from the original data lie} squarely in the middle of the posterior predictive distribution\edit{s} corresponding to the compositional model.
In sharp contrast, the simple model, which does not explicitly model inter-event dependencies, induces posterior predictive distributions over correlations that are concentrated away from and weaker than the actually observed correlations.


\subsection{Studying inter-event relationships}
\label{subsec:comp_interpretable}

\begin{figure}[ht]
  \centering
      \includegraphics[width=.9\textwidth]{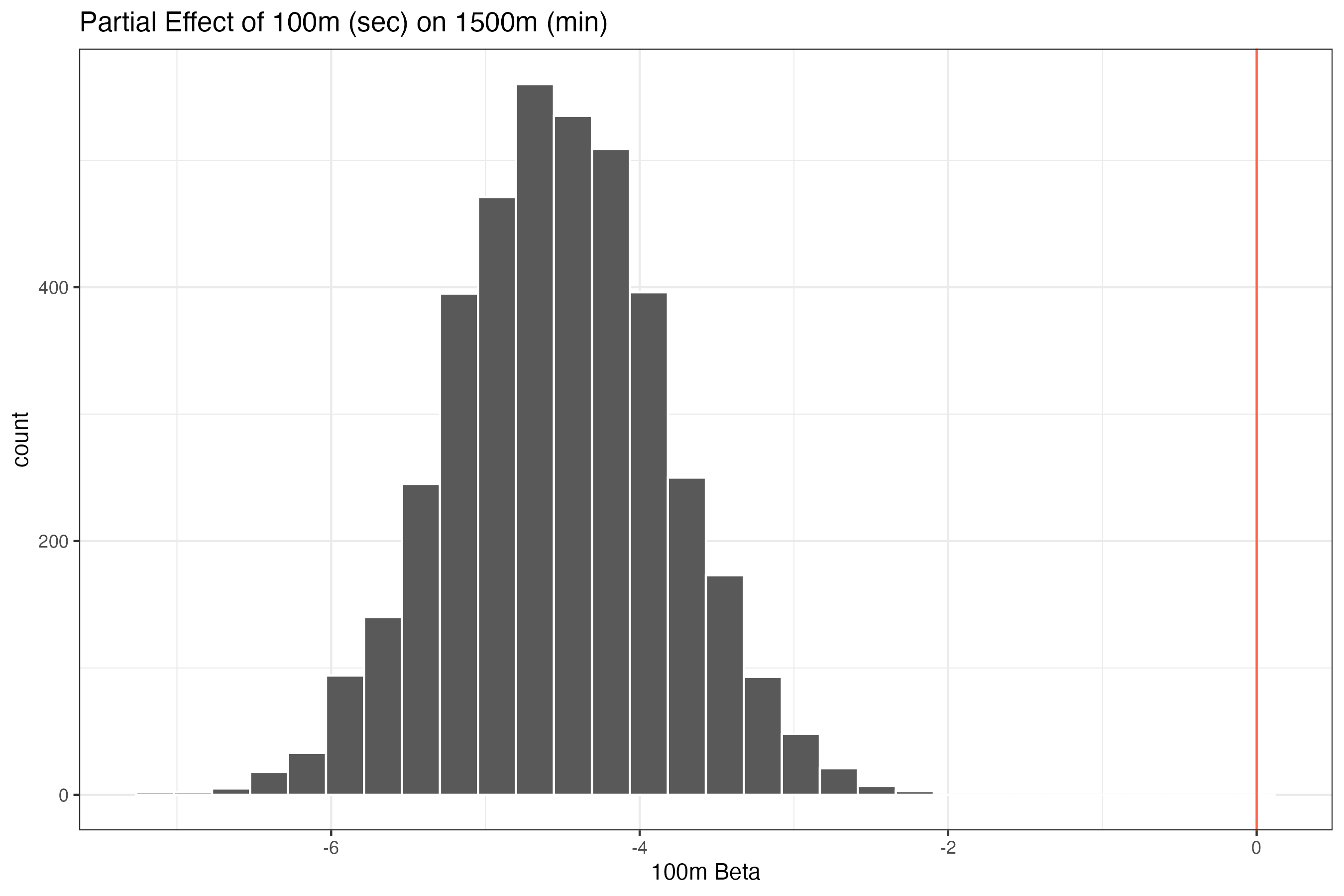}
     \caption{Histogram of posterior draws of $\beta$ associated with 100m for prediction of 1500m.}
     \label{fig:beta_hist} 
\end{figure}

To better understand the relationship between any two events, we can examine the posterior distributions of their modeled coefficient.
\Cref{fig:beta_hist} displays the distributions of the effect of the 100m sprint on 1500m performance. 
Our model estimates this relationship to be antagonistic. 
That is, an improvement of 0.1 second in the 100m, after accounting for age and the sequential effects of preceding events, is associated with an expected increase of .45 minutes in the 1500m. 

Before proceeding, we pause to stress that these associations are not necessarily causal.
A positive coefficient between the 100m and long jump may arise from shared underlying attributes or an emphasis on, say, \edit{anaerobic} training, rather than a direct transfer of skill from one event to another. 
This is particularly relevant given that our dataset consists only of top decathlon performances, where \edit{decathlete}s \edit{are typically} competitive across all events. \edit{\citet{stemmler2005detection} found that while all-rounders generally achieve higher scores than specialists, atheletes who excel in only one or two types of events (e.g., sprinting) can still achieve high scores.}
Consequently, these coefficients may be better interpreted as indicators of general performance patterns at the elite level rather than isolated event-on-event effects. 

\edit{\Cref{tab:coef_table} contains the posterior mean regression event coefficients for each decathlon discipline. For instance, the posterior mean coefficient corresponding to 100m, when modeling LJ, is -0.48. Positive coefficients whose 95\% credible interval do not contain 0 are bolded. Our results are comparable to previous grouping analyses of the decathlon. Similar to \citet{woolf2007grouping,schomaker2011model, cox2002analysis}, we find evidence of a grouping of `speed' events, namely 100m, LJ, SP, 400m, 110mH. We also find a grouping for `throwing' events - note that the coefficients highlighted in JT column include SP and DT, that the highlighted coefficient for the DT is SP, and that these coefficients are all positive. In contrast to the previous approaches, our model depicts 1500m with stronger relationships with 400m and JT.}

\begin{table}

  \caption{\edit{\label{tab:coef_table}Posterior mean regression coefficients by event. The $i,j$'th entry represents the coefficient corresponding to event $i$ when modeling event $j$. Positive coefficients whose 95\% credible interval do not contain 0 have been bolded.}}
  \centering
  \begin{tabular}[t]{llllllllll}
  \toprule
  Event & LJ & SP & HJ & 400m & 110mH & DT & PV & JT & 1500m\\
  \midrule
  100m & -0.48 & -0.14 & -0.09 & \textbf{{0.42}} & \textbf{{0.25}} & -0.03 & -0.07 & 0.00 & -0.09\\
  LJ & - & \textbf{{0.11}} & \textbf{{0.21}} & -0.11 & -0.09 & 0.00 & \textbf{0.05} & \textbf{0.05} & -0.03\\
  SP & - & - & \textbf{{0.12}} & -0.08 & -0.09 & \textbf{{0.36}} & \textbf{0.07} & \textbf{{0.19}} & \textbf{0.06}\\
  HJ & - & - & - & -0.05 & -0.07 & \textbf{0.03} & \textbf{0.05} & \textbf{0.04} & -0.06\\
  400m & - & - & - & - & \textbf{{0.21}} & 0.00 & -0.09 & -0.04 & \textbf{{0.44}}\\
  \addlinespace
  110mH & - & - & - & - & - & -0.08 & -0.12 & -0.03 & \textbf{0.03}\\
  DT & - & - & - & - & - & - & \textbf{{0.12}} & \textbf{{0.11}} & 0.00\\
  PV & - & - & - & - & - & - & - & \textbf{0.08} & -0.07\\
  JT & - & - & - & - & - & - & - & - & -0.12\\
  \bottomrule
  \end{tabular}
  \end{table}

\subsection{Simulating decathlon performance over time}
\label{subsec:sim_procedure}

We can forecast the entire arc of an individual decathlete's career with a posterior predictive simulation.
Specifically, for a given \edit{decathlete}, $i$, posterior sample, $s$, and age, we simulate a single decathlon performance as follows.
First, we simulate their 100m performance at that age by drawing
$$ 
Y^{\star(s)}_{i,100m} \sim \normaldist{\alpha_{i,100m}^{(s)} + \sum_{d=1}^{D}{\beta_{d,100m}^{(s)} \phi_{d}(\textrm{age})}}{\sigma_{100m}^{(s)}}.
$$
Then, having simulated their 100m performance time at that age, we can simulate their long jump performance by drawing 
$$ 
Y^{\star(s)}_{i,LJ} \sim \normaldist{\alpha_{i,LJ}^{(s)} + \sum_{d=1}^{D}{\beta_{d,LJ}^{(s)} \phi_{d}(\textrm{age})} + \gamma_{100m,LJ}^{(s)} Y_{i,100m}^{\star(s)}}{\sigma^{(s)}_{LJ}},
$$
where $Y_{i,100m}^{\star(s)}$ is the just-simulated 100m performance.
We then continue this process, simulating the performance in event $e$ using the corresponding posterior samples \emph{and} the simulated performances from the previous events:
 $$
Y_{i,e}^{(s)} \sim \normaldist{\alpha_{i,e}^{(s)} + \sum_{d=1}^{D}{\beta_{d,e}^{(s)} \phi_{d}(\textrm{age})} + \sum_{m = 1}^{e-1}{\gamma_{m,e}^{(s)} Y_{i,j,m}^{(s)}}}{\sigma_{e}^{(s)}}.
$$

At the end, we obtain a single sample from the posterior predictive distribution of performance in each event of a single decathlon, which we can then convert into a single overall score using \Cref{tab:point_params}.
Repeating this for every age in a fixed grid, and with every sample from the posterior distribution of the model parameters, yields a posterior distribution over each decathlete's age curve.
Importantly, our posterior predictive simulation strategy propagates and combines not only uncertainty about model parameters but the inherent variability in actual event performance. 

\begin{figure}
  \begin{subfigure}{.5\textwidth}
    \centering
    \includegraphics[width=\linewidth]{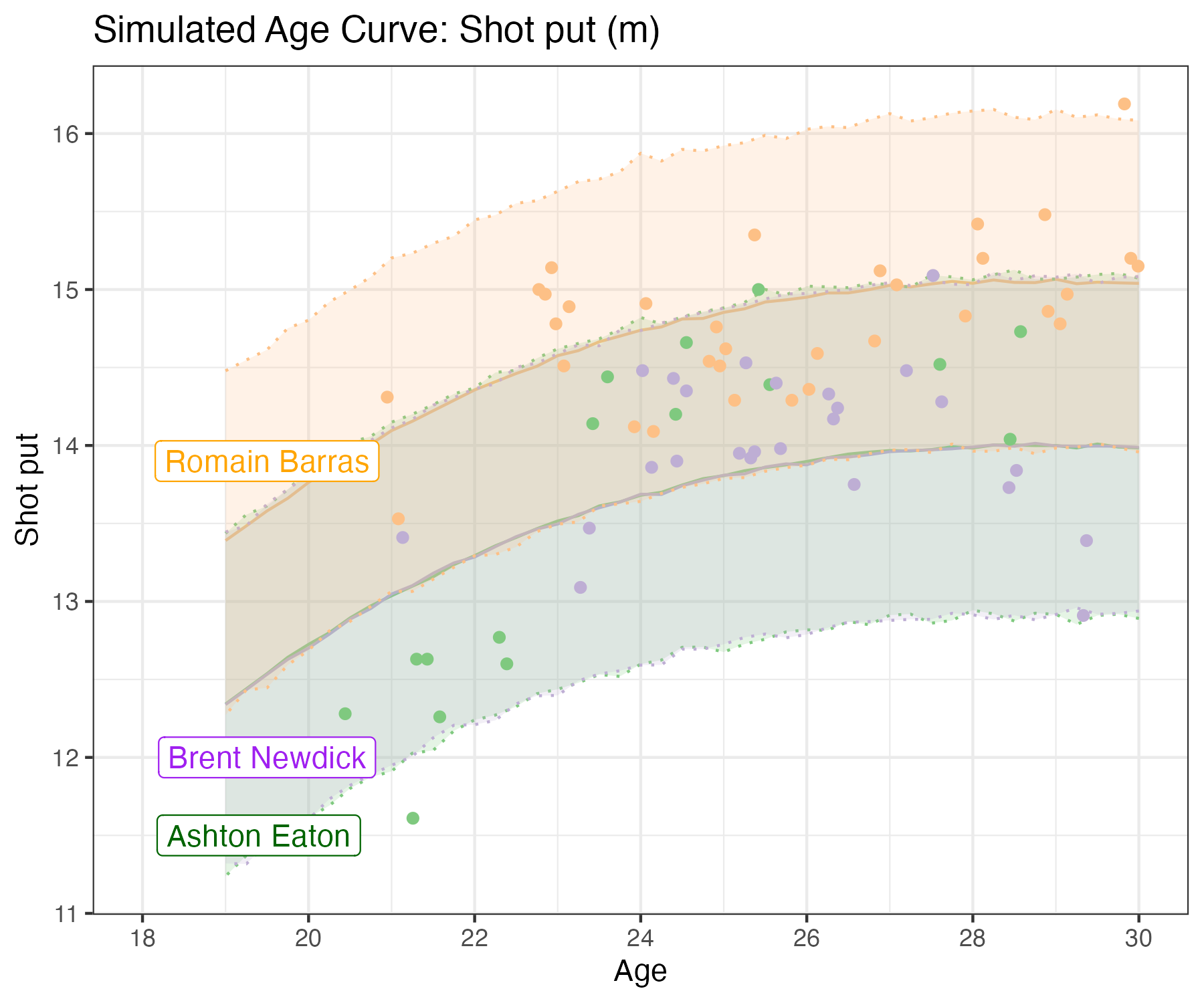}
    \caption{Shot Put}
  \end{subfigure}%
  \begin{subfigure}{.5\textwidth}
    \centering
    \includegraphics[width=\linewidth]{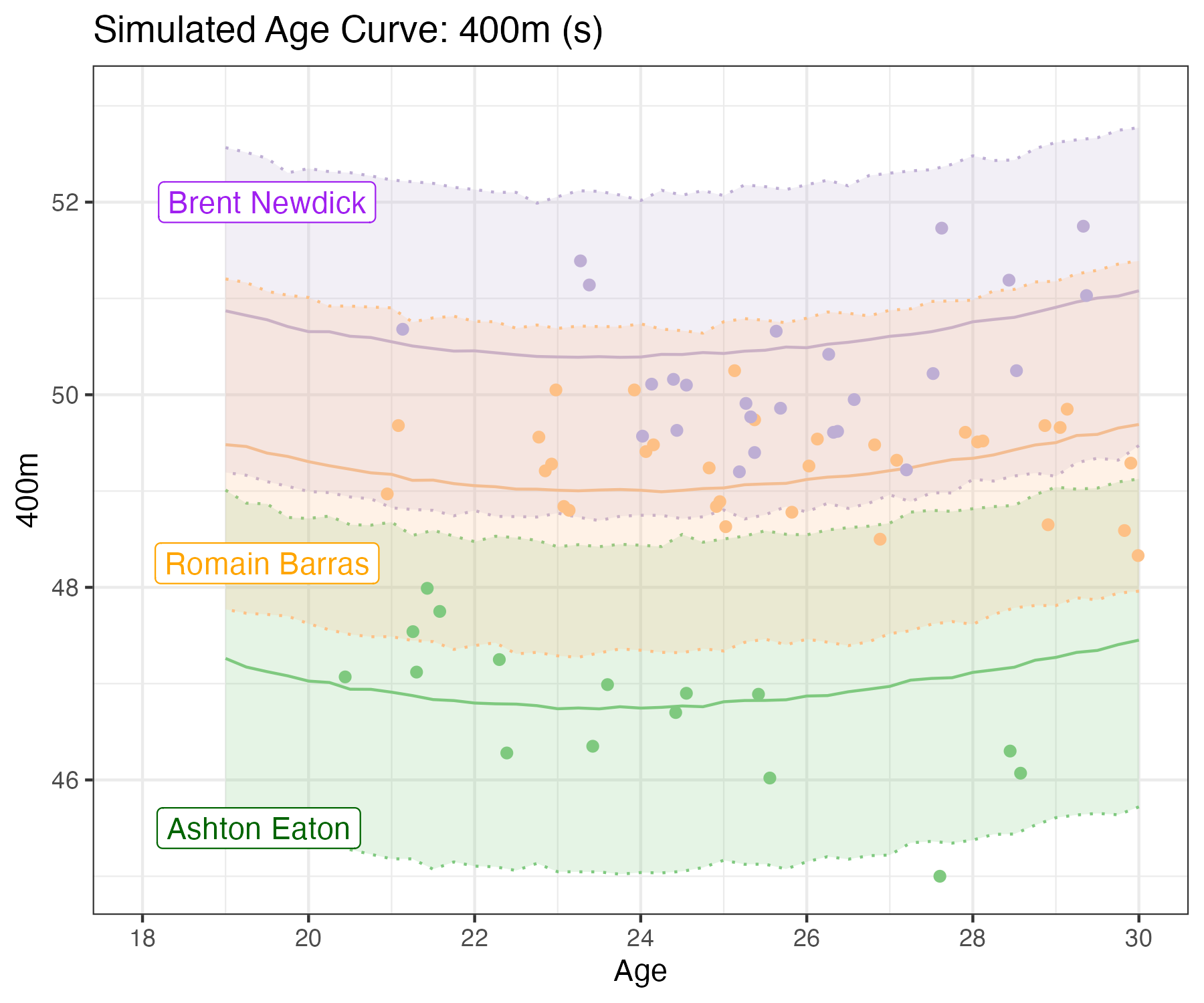}
    \caption{400m}
  \end{subfigure}
  \caption{Posterior predictive shotput (left) and 400m (right) intervals for selected \edit{decathlete}s. The shaded areas represent the 95\% posterior predictive interval for each respective model, and the solid line depicts the posterior mean. Each dot represents an observation for a given \edit{decathlete}. Greater distances for shot put and faster times for the 400m correspond to more points.}
    \label{fig:points_age_curve}
\end{figure}

\Cref{fig:points_age_curve} shows the point-wise posterior predictive mean and 95\% uncertainty intervals for several decathletes' age curves. 
Ashton Eaton, Romain Barras and Brendt Newdick are highlighted in \Cref{fig:points_age_curve} to compare progressions of decathletes with performances consistently across our age range. 
Reassuringly, the observed performances tend to lie within the point-wise 95\% predictive intervals. 
We also see benefits of modeling each event individually. 
The shapes of each age curve differ for different events. 
SP distances improve steadily over time while 400m times improve and then degrade over time.
Further, while \edit{decathlete}s display their peak SP distances in the later stages of their careers, their 400m times are predicted to peak in their early-20's. 

\subsection{Measuring latent skill}
\label{subsec:latent_skill}
Recall that our models assume that for each event, random \edit{decathlete} intercepts are drawn from a latent normal population with event-specific means and variances. 
This population, in some sense, captures the range of latent ability in each event, and the decathletes in our analysis can be viewed as a sample from the underlying population.

\begin{figure}[ht]
  \centering
    \includegraphics[width=.9\linewidth]{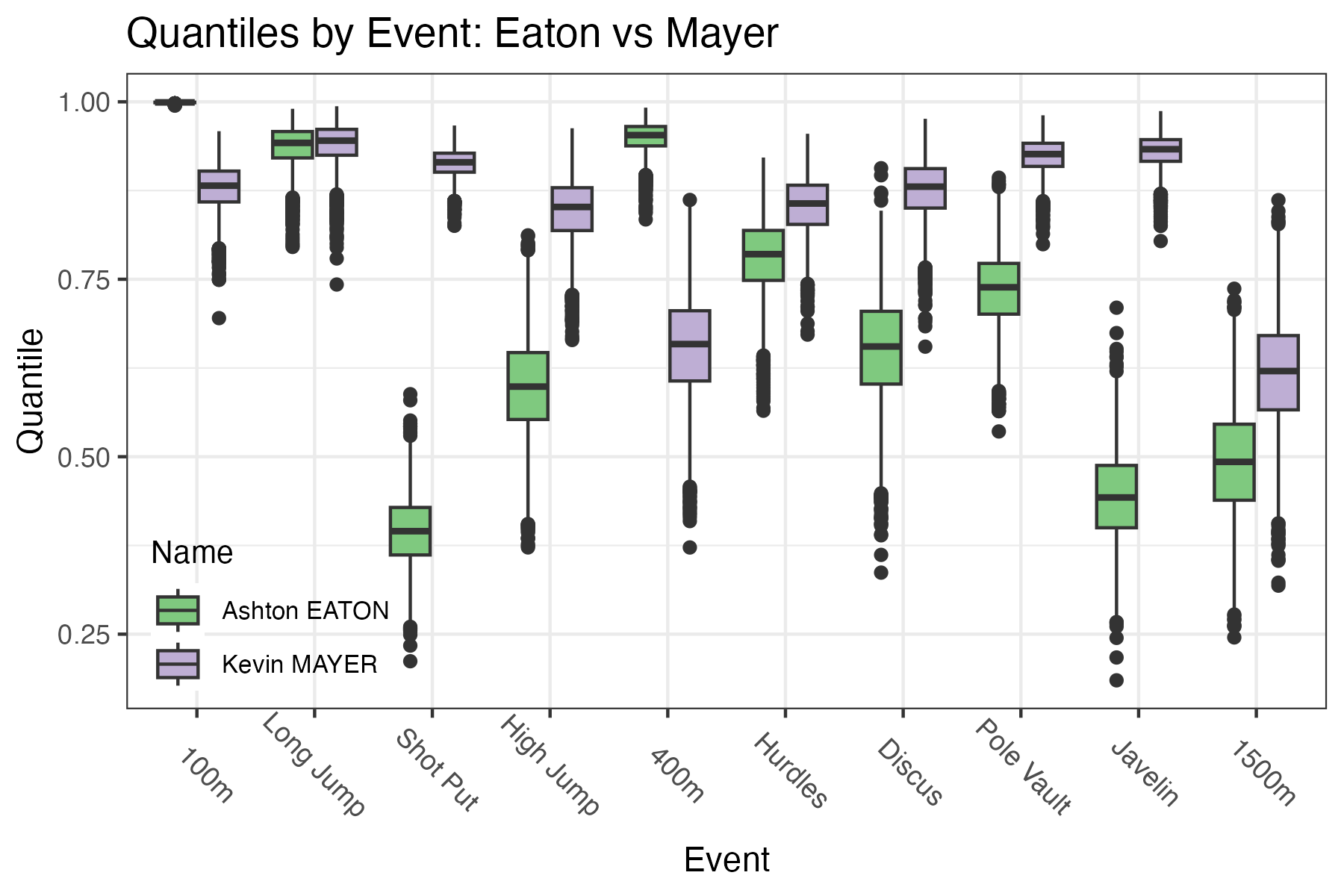}
    \caption{Boxplots of quantiles for event intercepts for Ashton Eaton and Kevin Mayer across 4000 MCMC samples.}
    \label{fig:intercepts_boxplot}
\end{figure}

To evaluate \edit{decathlete}-specific latent skill in an event, we use the posterior distribution of \edit{decathlete}-level random intercepts from our hierarchical model. Since slopes for age and other covariates are shared across \edit{decathlete}s, these intercepts capture persistent \edit{decathlete}-level differences in event performance. 
For any given \edit{decathlete} in our sample, we assess their latent skill in each event by computing the z-score and percentile associated with their event-intercept within the overall posterior distribution of \edit{decathlete} random intercepts. 
\Cref{fig:intercepts_boxplot} visualizes the posterior distribution of the Ashton Eaton's and Kevin Mayer's percentile within the latent population of skill in each discipline. 
We find that Eaton is among the elite at the 100m, LJ, and 400m, while Mayer is generally better than Eaton on the Day 2 events (hurdles, DT, PV, JT, and 1500m). 

\begin{figure}[ht]
    \centering
      \includegraphics[width=.85\linewidth]{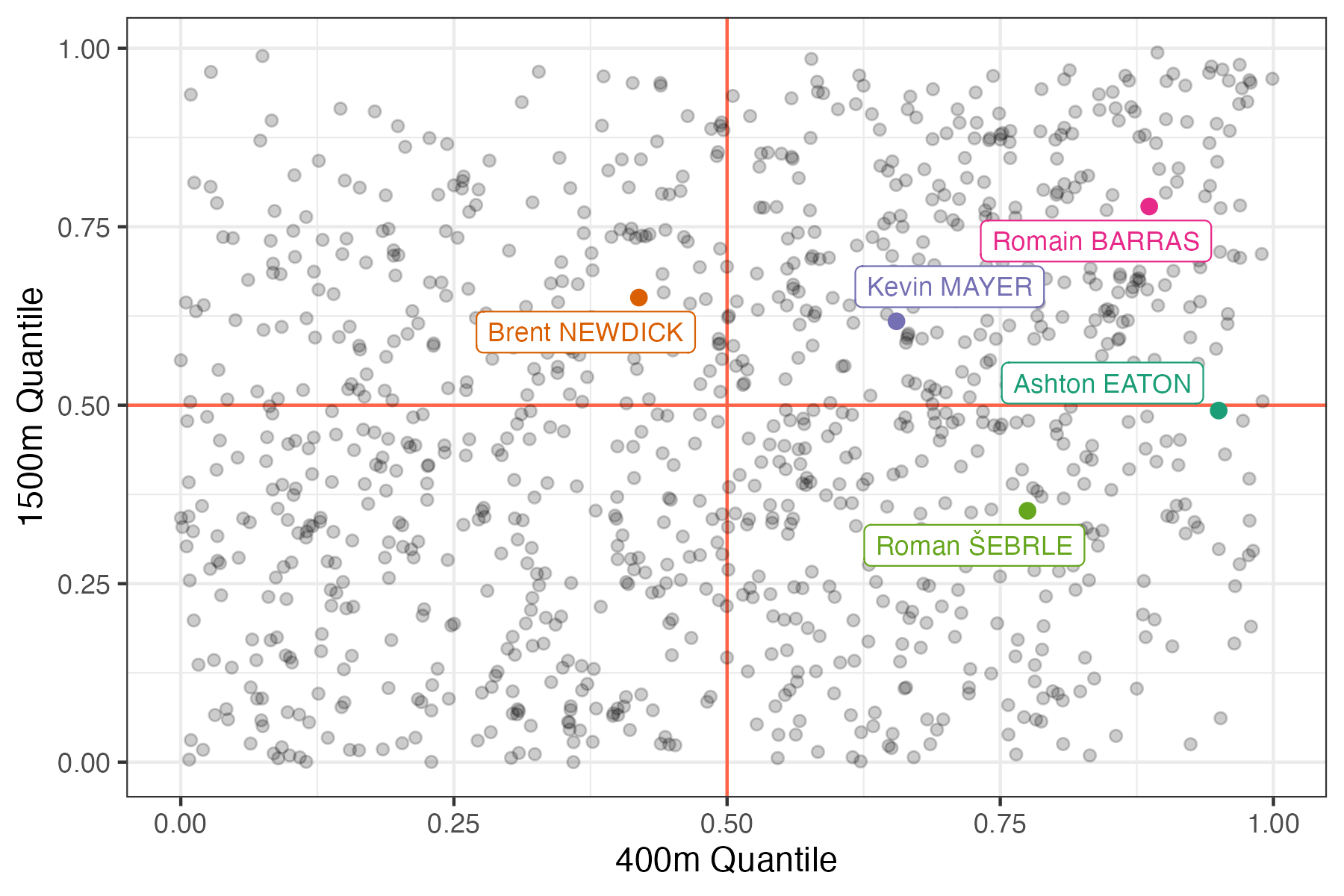}
      \caption{Mean posterior intercept percentiles in the 400m and 1500m for all \edit{decathlete}s, with selected \edit{decathlete}s highlighted. The 50th percentile for each event is marked with a red line.}
      \label{fig:intercepts_plot}
  \end{figure}

\Cref{fig:intercepts_plot} plots the posterior mean quantile for each decathlete's latent skill in the 400m and 1500m.
We find that Eaton ranks above the 90th-percentile in terms of his latent 400m skill but is in the middle-of-the-pack for the 1500m.
Mayer, by contrast, is above average in both events but not to a large degree.
Their examples suggest that achieving high decathlon scores does not require elite performance in all events.

\subsection{\edit{Decathlete} Profiles - Breaking 9200}
\label{subsec:profiles_sim}

In \Cref{subsec:latent_skill}, we calculated the mean posterior event-intercept percentiles in each event for selected \edit{decathlete}s in our sample.
We refer to this collection of posterior mean percentiles as a decathlete's \emph{profile}.
\Cref{tab:profile_table} shows the profiles for the five decathletes with the highest scores in our dataset. 
Given any profile (i.e., any combination of 10 quantiles), we can simulate a career's worth of decathlons for a hypothetical decathlete with that profile using a procedure very similar to the one described in \Cref{subsec:sim_procedure}. 
Specifically, for each posterior sample $s,$ we set the random intercept $\alpha_{e}^{s}$ to the corresponding quantile of the $\normaldist{\mu^{(s)}_{e}}{\sigma^{2(s)}_{\alpha,e}}$ distribution. 
Then, we simulate 2 decathlons for every year between the ages 19 and 30, and compute the proportion of simulations where this \edit{decathlete} breaks 9200 points.

For instance, to simulate the career of a decathlete with Asthon Eaton's profile --- that is, someone who is elite in the 100m, long jump, and 400m but slightly below average in the shot put and javelin toss --- we would sample the the 99.9\%-quantile for the 100m intercept, the 94\%-quantile for the LJ, and so on. 
We find that \edit{a decathlete}\edit{, over the course of their entire career,} with Eaton's profile broke 9200 points in just 6\% of our posterior samples while decathletes with Kevin Mayer's and Roman \v{S}ebrle's profiles respectively break 9200 with posterior probabilities of 7\% and 10\%\edit{, assuming \edit{decathlete}s compete consistently for this period of time.  We note that these simulations likely overestimate these probabilities, as they do not account for injury, mental fatigue, recovery time, or pressure. Additionally, our model and results are based on historical data, and do not account for unexpected breakthroughs that will systematically change human performance (e.g., new track surfaces, novel performance enhancement supplements, etc.)}. 

We repeat this procedure for the various synthetic \edit{decathlete} profiles. 
We consider \edit{a decathlete} that specializes in the Day 1 events, with 95th percentile intercepts in the first five events and 50th percentile intercepts in the last five events. 
This \edit{decathlete} breaks 9200 points in only 2\% of our simulations. 
The analogous Day 2 specialist never broke 9200 in our simulations. 

\begin{figure}
  \centering
    \includegraphics[width=.9\linewidth]{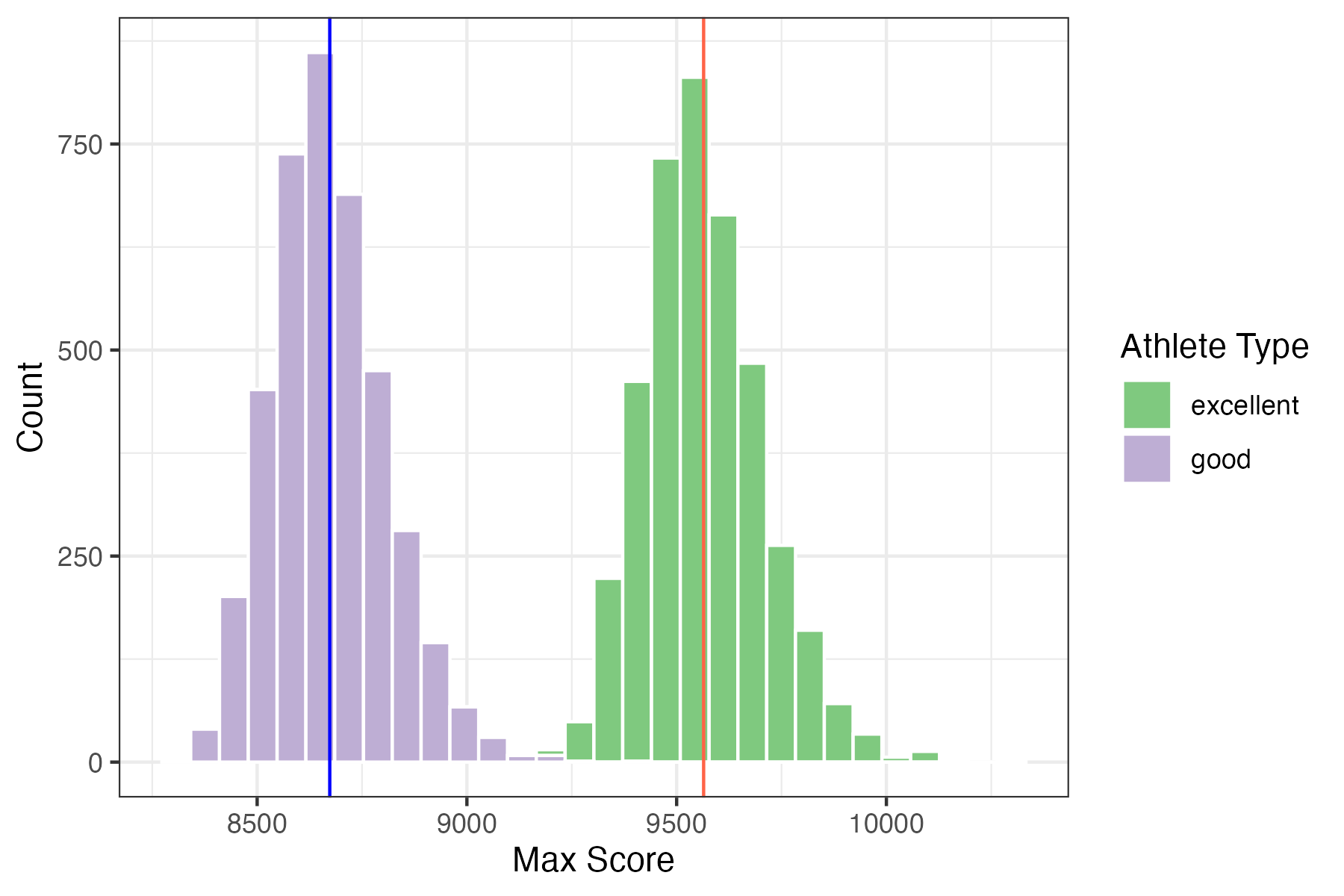}
    \caption{Histogram of greatest decathlon performances for synthetic \edit{decathlete}s, excellent (green) and good (purple), across 4000 simulated careers. The vertical line depicts the average highest score for each synthetic \edit{decathlete}.}
    \label{fig:synth_hist}
\end{figure}

Perhaps unsurprisingly, an unrealistic ``unicorn'' \edit{decathlete}, who is consistently 95th percentile across all ten events, does break the 9200 mark in virtually all of our simulations.
But the somewhat more realistic ``good'' decathlete, who consistently performs at the 80th percentile across all events, almost never breaks 9200. 
\Cref{fig:synth_hist} shows histograms for the highest scores in each simulated career for these synthetic \edit{decathlete}s. 
There is minimal overlap for their max scores at their respective skill levels, with the ``unicorn'' \edit{decathlete}'s average high score being 9561 points and the ``good'' \edit{decathlete}'s average high score at 8671.
So, returning to our original question, we conclude that breaking 9200, while technically feasible, is highly unlikely and would be an extremely impressive feat. 

\begin{table}
  \caption{\label{tab:profile_table}Profiles of last world record holders and synthetic \edit{decathlete}s. Entries consist of each event's intercept percentile. We calculate the proportion of simulated careers in which \edit{a decathlete} with the given profile breaks 9200 points.}
  \centering
  \begin{tabular}{lrrrrrrrrrrl} 
  \toprule
  Name & 100m & LJ & SP & HJ & 400m & 110mH & DT & PV & JT & 1500m & 9200?\\
  \midrule
  K. Mayer & 0.88 & 0.94 & 0.91 & 0.85 & 0.66 & 0.85 & 0.88 & 0.92 & 0.93 & 0.62 & 7.7\\
  A. Eaton & 1.00 & 0.94 & 0.40 & 0.60 & 0.95 & 0.78 & 0.65 & 0.74 & 0.44 & 0.49 & 5.65\\
  R. \v{S}ebrle & 0.91 & 1.00 & 0.92 & 0.94 & 0.77 & 0.55 & 0.79 & 0.57 & 0.95 & 0.35 & 10.325\\
  D. Warner & 1.00 & 0.71 & 0.52 & 0.67 & 0.57 & 0.90 & 0.82 & 0.26 & 0.77 & 0.44 & 1.825\\
  R. Dvo\v{r}\'{a}k & 0.92 & 0.95 & 0.97 & 0.55 & 0.69 & 0.74 & 0.44 & 0.34 & 0.96 & 0.67 & 0.6\\
  \addlinespace
  Day 1 & 0.95 & 0.95 & 0.95 & 0.95 & 0.95 & 0.50 & 0.50 & 0.50 & 0.50 & 0.50 & 2.4\\
  Day 2 & 0.50 & 0.50 & 0.50 & 0.50 & 0.50 & 0.95 & 0.95 & 0.95 & 0.95 & 0.95 & 0\\
  Excellent & 0.95 & 0.95 & 0.95 & 0.95 & 0.95 & 0.95 & 0.95 & 0.95 & 0.95 & 0.95 & 99.95\\
  Good & 0.80 & 0.80 & 0.80 & 0.80 & 0.80 & 0.80 & 0.80 & 0.80 & 0.80 & 0.80 & 0.075\\
  \bottomrule
  \end{tabular}
\end{table}

%% file: discussion.tex
In this paper, we developed a new compositional Bayesian model that can predict overall decathlon scores and individual discipline performance over the course of a decathlete's career.
We showed that our compositional model better captured inter-event dependencies than simpler alternatives.
Using our model, we introduced a profile for each decathlete, which summarizes their latent skill in an easily-interpretable way.
Examining these profiles revealed that elite decathlon performance tend not to arise from elite performance across all ten disciplines.
Rather, the top decathletes tend to be elite at only a few events (e.g., Ashton Eaton) or good but not elite in many (e.g., Kevin Mayer).
Based on posterior predictive simulations of several synthetic decathlete profiles, we concluded that breaking the 9200 point limit is \edit{possible, but} unlikely.

There are several potential extensions to our overall modeling.
First, we were unable to account for important variables that affect decathlon performance like temperature, wind speeds, and shoe and surface technology.
That said, it is conceptually straightforward to elaborate the model in \Cref{eq:compositional_model} to include additional covariates.
More substantively, we fit each event-specific model separately, which renders the event-specific intercepts for a given \edit{decathlete} independent \textit{a posteriori}.
A better way to capture \emph{latent} inter-event dependence would be to fit a single model for all event performances in which the vector of \edit{decathlete}-specific random intercepts was drawn from a normal population with non-diagonal covariance matrix. 
Additionally, as noted by \citet{schuckers2023estimation}, our analysis may be subject to a certain amount of selection bias, as only the most successful decathletes are observed at the extremes of the age range considered. 
While this is certainly of concern, we believe that the selection bias is less severe in track and field than in professional sports like baseball or hockey.
Finally, our compositional model only allowed for linear dependence between events.
We leave development and investigation of a non-linear compositional model to future work.

%% file: acknowledgement.tex
The authors thank Tristan Faure for his early contributions to the original conception of this project and for assistance with initial data collection during its preliminary phase.

%% file: appendix.tex
\subsection{Point calculation details}
The formula for calculating points for each individual event is given below:
\begin{equation}
    \label{eq:point_calc}
    \text{Points} = 
    \begin{cases}

        a \cdot (b-y)^c& \text{if track event} \\
    
        a \cdot (y-b)^c & \text{if field event}
    
    \end{cases}
\end{equation}
where $a$, $b$, and $c$ are given by \Cref{tab:point_params}, and $y$ is the \edit{decathlete}'s time, distance, or height. Faster times in track events (100m, 400m, hurdles, and 1500m) and greater distances and heights in field events produce higher scores.

\begin{table}
    \caption{\label{tab:point_params}Parameter values by event for point calculation }
    \centering
    \begin{tabular}[t]{lrrr}
    \toprule
    Event & a & b & c\\
    \midrule
    100 m & 25.435 & 18.0 & 1.81\\
    Long jump & 0.144 & 220.0 & 1.40\\
    Shot put & 51.390 & 1.5 & 1.05\\
    High jump & 0.847 & 75.0 & 1.42\\
    400 m & 1.538 & 82.0 & 1.81\\
    \addlinespace
    110 m hurdles & 5.744 & 28.5 & 1.92\\
    Discus throw & 12.910 & 4.0 & 1.10\\
    Pole vault & 0.280 & 100.0 & 1.35\\
    Javelin throw & 10.140 & 7.0 & 1.08\\
    1500 m & 0.038 & 480.0 & 1.85\\
    \bottomrule
    \end{tabular}
\end{table}

\subsection{Model comparisons with real data}
We compared the out of sample accuracy on real decathlon data for each model under two frameworks. 
The first framework is a `general' case, where we created ten 90\%-10\% training/testing splits, with observations placed in the test split at random. 
In `tail' framework, we create ten training/testing splits by selecting 10\% of the decathletes and using their last decathlon observation as the test set, and all other observations in the training dataset. 
\Cref{tab:gen_table,tab:future_table} contain the results (SMSE) from the real decathlon data bakeoff experiments with randomly removed observations and tail observations. 



\begin{sidewaystable}

    \caption{\label{tab:gen_table}Mean standardized MSE across 10 cross validations in predicting decathlon performance with randomly removed observations. }
    \centering
    \begin{tabular}[t]{llrrrrrrrrrrr}
    \toprule
    model & basis & 100m & LJ & SP & HJ & 400m & 110mH & DT & PV & \edit{JT} & 1500m & points\\
    \midrule
    baseline & cubic & - & - & - & - & - & - & - & - & - & - & 0.234\\
    simple & cubic & 0.309 & 0.414 & 0.188 & 0.343 & 0.329 & 0.311 & 0.262 & 0.294 & 0.295 & 0.413 & 0.235\\
    compositional & cubic & 0.309 & 0.414 & 0.188 & 0.344 & 0.329 & 0.312 & 0.262 & 0.294 & 0.294 & 0.414 & 0.235\\
    \midrule
    baseline & spline & - & - & - & - & - & - & - & - & - & - & 0.235\\
    simple & spline & 0.309 & 0.414 & 0.188 & 0.344 & 0.329 & 0.311 & 0.263 & 0.294 & 0.295 & 0.414 & 0.235\\
    compositional & spline & 0.309 & 0.414 & 0.188 & 0.344 & 0.329 & 0.312 & 0.262 & 0.294 & 0.294 & 0.414 & 0.235\\
    \bottomrule
    \end{tabular}
\end{sidewaystable}

\begin{sidewaystable}

    \caption{\label{tab:future_table} Mean standardized MSE across 10 cross validations tail-removed observations for decathletes. }
    \centering
    \begin{tabular}[t]{llrrrrrrrrrrr}
    \toprule
    model and prior & basis & 100m & LJ & SP & HJ & 400m & 110mH & DT & PV & \edit{JT} & 1500m & points\\
    \midrule
    baseline none & cubic & - & - & - & - & - & - & - & - & - & - & 0.358\\
    simple none & cubic & 0.337 & 0.471 & 0.238 & 0.414 & 0.406 & 0.447 & 0.301 & 0.372 & 0.315 & 0.482 & 0.362\\
    compositional none & cubic & 0.337 & 0.469 & 0.239 & 0.415 & 0.407 & 0.445 & 0.300 & 0.371 & 0.315 & 0.482 & 0.362\\
    \midrule
    baseline & spline & - & - & - & - & - & - & - & - & - & - & 0.359\\
    simple & spline & 0.338 & 0.470 & 0.238 & 0.415 & 0.409 & 0.448 & 0.301 & 0.371 & 0.317 & 0.484 & 0.363\\
    compositional & spline & 0.337 & 0.469 & 0.239 & 0.415 & 0.408 & 0.445 & 0.300 & 0.369 & 0.315 & 0.482 & 0.363\\
    \bottomrule
    \end{tabular}
    \end{sidewaystable}

\subsection{Parameter recovery with known \texorpdfstring{$\bm{\beta}$}{TEXT}}
\label{subsec:param_exp}

\Cref{tab:prop_table_cube_comp} displays the proportion of 95\% credible intervals containing the true parameter associated with the inter-event relationship over 200 simulated datasets, as described in \Cref{sec:experiments}. We specified a known structure of linear dependence between events and estimated coefficients for the age polynomials and preceding events using the original decathlon data.
We then simulated 200 datasets with 8668 observations.
After fitting the cubic, compositional model to the 200 simulated datasets, we studied the posterior distributions over the $\beta$ coefficients corresponding to age and preceding events. 
For each coefficient, we computed the proportion of 95\% posterior intervals containing the true value used in simulation.

\begin{table}
    \caption{\label{tab:prop_table_cube_comp}Proportion of 95\% credible intervals containing the true parameter associated with corresponding predictor over 200 simulations. Entries are rounded to the second digit.}
    \centering
    \begin{tabular}[t]{lrrrrrrrrrr}
    \toprule
    predictor & 100m & LJ & SP & HJ & 400m & 110mH & DT & PV & \edit{JT} & 1500m\\
    \midrule
    age & 0.95 & 0.96 & 0.96 & 0.96 & 0.94 & 0.95 & 0.96 & 0.96 & 0.92 & 0.95\\
    $\text{age}^2$ & 0.94 & 0.94 & 0.96 & 0.92 & 0.96 & 0.96 & 0.95 & 0.96 & 0.96 & 0.98\\
    $\text{age}^3$  & 0.92 & 0.96 & 0.94 & 0.94 & 0.96 & 0.94 & 0.95 & 0.96 & 0.93 & 0.97\\
    100m & - & 0.97 & 0.94 & 0.98 & 0.97 & 0.93 & 0.96 & 0.98 & 0.95 & 0.96\\
    LJ & - & - & 0.97 & 0.94 & 0.96 & 0.93 & 0.94 & 0.95 & 0.94 & 0.96\\
    \addlinespace
    SP & - & - & - & 0.98 & 0.96 & 0.94 & 0.98 & 0.96 & 0.96 & 0.94\\
    HJ & - & - & - & - & 0.97 & 0.94 & 0.96 & 0.96 & 0.96 & 0.98\\
    400m & - & - & - & - & - & 0.93 & 0.95 & 0.96 & 0.96 & 0.96\\
    110mH & - & - & - & - & - & - & 0.95 & 0.93 & 0.92 & 0.96\\
    DT & - & - & - & - & - & - & - & 0.96 & 0.95 & 0.96\\
    \addlinespace
    PV & - & - & - & - & - & - & - & - & 0.90 & 0.94\\
    JT & - & - & - & - & - & - & - & - & - & 0.94\\
    \bottomrule
    \end{tabular}
\end{table}


\subsection{Empirical and posterior predictive correlations between decathlon events}
\label{subsec:cor_experiment}

\Cref{tab:emp_cor} contains the empirical correlations between the decathlon events from the observed data. \Cref{tab:sim_cor_simple,tab:sim_cor_comp} contain the 2.5\% and 97.5\% quantiles from the posterior predictive correlations from the simulated datasets generated by the simple and compositional models respectively, as described in \Cref{sec:experiments}.

\begin{table}

    \caption{\label{tab:emp_cor}Empirical correlations between decathlon events in observed data, rounded to 2 digits.}
    \centering
    \begin{tabular}[t]{lrrrrrrrrrr}
    \toprule
    Event & 100m & LJ & SP & HJ & 400m & 110mH & DT & PV & JT & 1500m\\
    \midrule
    100m & 1.00 & -0.54 & -0.20 & -0.20 & 0.66 & 0.52 & -0.18 & -0.25 & -0.13 & 0.10\\
    LJ & - & 1.00 & 0.33 & 0.43 & -0.45 & -0.47 & 0.29 & 0.34 & 0.24 & -0.14\\
    SP & - & - & 1.00 & 0.31 & -0.15 & -0.36 & 0.73 & 0.38 & 0.51 & -0.01\\
    HJ & - & - & - & 1.00 & -0.21 & -0.34 & 0.28 & 0.29 & 0.22 & -0.10\\
    400m & - & - & - & - & 1.00 & 0.46 & -0.13 & -0.25 & -0.11 & 0.46\\
    \addlinespace
    110mH & - & - & - & - & - & 1.00 & -0.32 & -0.37 & -0.24 & 0.12\\
    DT & - & - & - & - & - & - & 1.00 & 0.40 & 0.48 & -0.02\\
    PV & - & - & - & - & - & - & - & 1.00 & 0.31 & -0.20\\
    JT & - & - & - & - & - & - & - & - & 1.00 & -0.09\\
    1500m & - & - & - & - & - & - & - & - & - & 1.00\\
    \bottomrule
    \end{tabular}
    \end{table}

\begin{sidewaystable}
        \caption{\label{tab:sim_cor_simple}2.5\% and 97.5\% quantiles for posterior predictive correlation between decathlon events from 2000 simulated datasets from the simple model, rounded to 2 digits.}
        \centering
        \begin{tabular}[t]{llllllllll}
        \toprule
        Event & LJ & SP & HJ & 400m & 110mH & DT & PV & JT & 1500m\\
        \midrule
        100m & -0.41, -0.37 & -0.16, -0.13 & -0.16, -0.12 & 0.49, 0.52 & 0.37, 0.41 & -0.15, -0.12 & -0.19, -0.16 & -0.11, -0.07 & 0.03, 0.07\\
        LJ & 1, 1 & 0.24, 0.28 & 0.29, 0.33 & -0.35, -0.31 & -0.37, -0.33 & 0.23, 0.26 & 0.24, 0.28 & 0.16, 0.2 & -0.1, -0.06\\
        SP & - & 1, 1 & 0.25, 0.28 & -0.12, -0.08 & -0.33, -0.29 & 0.64, 0.67 & 0.32, 0.35 & 0.44, 0.47 & -0.01, 0.03\\
        HJ & - & - & 1, 1 & -0.17, -0.13 & -0.29, -0.25 & 0.23, 0.26 & 0.22, 0.26 & 0.17, 0.2 & -0.07, -0.03\\
        400m & - & - & - & 1, 1 & 0.31, 0.35 & -0.11, -0.07 & -0.2, -0.16 & -0.09, -0.05 & 0.28, 0.32\\
        \addlinespace
        110mH & - & - & - & - & 1, 1 & -0.29, -0.25 & -0.31, -0.27 & -0.21, -0.18 & 0.03, 0.07\\
        DT & - & - & - & - & - & 1, 1 & 0.32, 0.36 & 0.41, 0.44 & -0.02, 0.02\\
        PV & - & - & - & - & - & - & 1, 1 & 0.24, 0.28 & -0.16, -0.12\\
        JT & - & - & - & - & - & - & - & 1, 1 & -0.06, -0.02\\
        1500m & - & - & - & - & - & - & - & - & 1, 1\\
        \bottomrule
        \end{tabular}

\end{sidewaystable}

\begin{sidewaystable}

    \caption{\label{tab:sim_cor_comp}2.5\% and 97.5\% quantiles for posterior predictive correlation between decathlon events from 2000 simulated datasets from the compositional model, rounded to 2 digits.}
    \centering
    \begin{tabular}[t]{llllllllll}
    \toprule
    Event & LJ & SP & HJ & 400m & 110mH & DT & PV & JT & 1500m\\
    \midrule
    100m & -0.55, -0.52 & -0.22, -0.18 & -0.22, -0.18 & 0.64, 0.67 & 0.5, 0.54 & -0.19, -0.16 & -0.27, -0.23 & -0.15, -0.11 & 0.09, 0.13\\
    LJ & 1, 1 & 0.3, 0.34 & 0.39, 0.43 & -0.47, -0.43 & -0.48, -0.45 & 0.26, 0.3 & 0.31, 0.35 & 0.21, 0.25 & -0.17, -0.12\\
    SP & - & 1, 1 & 0.29, 0.32 & -0.17, -0.13 & -0.37, -0.34 & 0.71, 0.73 & 0.36, 0.39 & 0.49, 0.52 & -0.03, 0.01\\
    HJ & - & - & 1, 1 & -0.23, -0.19 & -0.35, -0.31 & 0.25, 0.29 & 0.26, 0.3 & 0.2, 0.24 & -0.13, -0.08\\
    400m & - & - & - & 1, 1 & 0.44, 0.48 & -0.15, -0.11 & -0.27, -0.24 & -0.14, -0.1 & 0.43, 0.47\\
    \addlinespace
    110mH & - & - & - & - & 1, 1 & -0.33, -0.3 & -0.38, -0.35 & -0.26, -0.22 & 0.1, 0.15\\
    DT & - & - & - & - & - & 1, 1 & 0.37, 0.41 & 0.46, 0.49 & -0.04, 0\\
    PV & - & - & - & - & - & - & 1, 1 & 0.28, 0.32 & -0.22, -0.17\\
    JT & - & - & - & - & - & - & - & 1, 1 & -0.12, -0.08\\
    1500m & - & - & - & - & - & - & - & - & 1, 1\\
    \bottomrule
    \end{tabular}
    \end{sidewaystable}